\documentclass[letterpaper,11pt]{emulateapj}
\usepackage{graphicx}
\usepackage{color}
\usepackage{dashrule}
\usepackage{multirow,bigstrut,bigdelim}
\usepackage{epsfig}

\usepackage{graphics}
\usepackage{graphicx}
\begin{document}
\title{The extremely high peak energy of GRB 110721A in the context of a
dissipative photosphere synchrotron emission model}
\author{P\'eter Veres\altaffilmark{1,2,3,*},  Bin-Bin Zhang\altaffilmark{1}, P\'eter \Mesz\altaffilmark{1,2,3}}
\altaffiltext{1}{Department of Astronomy and Astrophysics, Pennsylvania State
University, 525 Davey Lab, University Park, PA 16802, USA}
\altaffiltext{2}{Department of Physics, Pennsylvania State University,
University Park, PA 16802, USA}
\altaffiltext{3}{Center for Particle Astrophysics, Pennsylvania State
University, University Park, PA 16802, USA}
\altaffiltext{*}{Email: veresp@psu.edu}
\date{\today} 
\def\ve{\varepsilon}
\def\gbm{{\it GBM }}
\def\lat{{\it LAT }}
\def\fermi{{\it Fermi }}
\newcommand{\Mesz}{{M\'esz\'aros}}
\def\mathnew{\mathsurround=0pt}
\def\simov#1#2{\lower .5pt\vbox{\baselineskip0pt \lineskip-.5pt
      \ialign{$\mathnew#1\hfil##\hfil$\crcr#2\crcr\sim\crcr}}}
\def\simg{\mathrel{\mathpalette\simov >}}
\def\siml{\mathrel{\mathpalette\simov <}}
\def\beq{\begin{equation}}
\def\enq{\end{equation}}
\def\bea{\begin{eqnarray}}
\def\ena{\end{eqnarray}}
\def\bitm{\bibitem}
\def\msun{M_\odot}
\def\L54{L_{54}}
\def\E55{E_{55}}
\def\et3{\eta_3}
\def\th1{\theta_{-1}}
\def\r07{r_{0,7}}
\def\x05{x_{0.5}}
\def\et600{\eta_{600}}
\def\et3{\eta_3}
\def\rph{r_{ph}}
\def\vareps{\varepsilon}
\def\fflunit{\hbox{~erg cm}^{-2}~\hbox{s}^{-1}}
\def\eps{\epsilon}
\def\ve{\varepsilon}
\def\muh{\hat{\mu}}
\def\cm{\hbox{~cm}}
\def\s{\hbox{~s}}
\def\gev{\hbox{~GeV}}
\def\GeV{\hbox{~GeV}}
\def\MeV{\hbox{~MeV}}
\def\kev{\hbox{~keV}}
\def\keV{\hbox{~keV}}
\def\eV{\hbox{~eV}}
\def\erg{\hbox{~erg}}
\def\s{{\hbox{~s}}
\def\cm2{\hbox{~cm}^2}}
\def\para{\parallel}
\defcitealias{Veres+12magnetic}{VM12}
\begin{abstract}
The Fermi observations of GRB 110721A \citep{Fermi+12epeak} have revealed an
unusually high peak energy $\sim 15$ MeV in the first time bin of the prompt
emission. We find that an interpretation is unlikely in terms of internal shock
models, and confirm that a standard black-body photospheric model also falls
short. We show that dissipative photospheric synchrotron models  ranging
from extreme magnetically dominated to baryon dominated dynamics, on the other
hand, are able to accommodate such high peak values.
\end{abstract}

\section{Introduction}
\label{sec:intro}
High energy ($>$ 100 MeV) observations of gamma-ray bursts (GRBs) have gained
renewed interest since the launch of Fermi LAT \citep{Atwood+09LAT}.  This
instrument has revealed a curious diversity in the high-energy behavior of GRBs
and consequently represents a new challenge for models  \citep[see ][ for a
recent review]{Mesz12GRBCTA}.  For GRB 110726A \citep{Fermi+12epeak},	the
extremely high $\ve_{\rm peak}\approx 15 \MeV$ peak energy (peak of the $\ve
F_\ve$ spectrum),   which was observed right after the onset of the burst,
makes its interpretation in the framework of a simple internal shock
synchrotron model challenging. Interpreting the peak as a modified black-body
from a simple photosphere is also difficult, as it can only account for
energies up to few MeVs, and even the subsequent time-bins show  peak fluxes
which, even though lower, are still unusually high. Such black-body photosphere
models	have been considered by \citet{Fan+12corr} to explain a
$L\propto\ve_{\rm peak}^{0.4}$ correlation found  by
\citet{Ghirlanda+12comoving},  but as shown by \citet{B-Zhang+12}, they must
lie below a line in the $\ve_{\rm peak} -L$ plane which excludes the observed
values for GRB 110721A.

Here we show that the high peak energy and luminosity of GRB110721A can be
interpreted in the framework of a different class of photospheres, where
dissipation occurs near the photospheric radius, and the spectrum is
characterized by a synchrotron peak. Dissipative synchrotron photospheres have
been considered for baryonic dominated regimes by, e.g.  \citet{Rees+05photdis,
Peer+06phot, Beloborodov10phot} and for magnetically dominated regimes by
\citet{Giannios06flare, Giannios12peak}.	Here we
consider such photospheres for an arbitrary acceleration law, characteristic of
either a baryonic or a magnetic dominated regime.  Also the observed thermal
bump close to $100 \kev$ is naturally incorporated in this model.

\section{Dynamics, photospheres and emission regions}
\label{sec:dyn}

The initial acceleration behavior of the jet material is taken to be given by
\begin{equation}
{\Gamma(r)}\propto \left\{
\begin{array}{lll}
r^{\mu}		&	{\rm if}	& r<r_{\rm sat}\\
{\rm const.}	&	{\rm if}	& r_{\rm sat}<r,
\end{array}
\right.
\label{eq:accel}
\end{equation}
where $r_{\rm sat}$ is the saturation radius beyond which the flow reaches its
asymptotic coasting value $\eta=\langle L\rangle/ \langle{\dot M}c^2 \rangle
=\Gamma_{\rm final}$.  Based on the analysis of \citep[e.g.][]{Drenkhahn02,
Meszaros+11gevmag}, for simple conical magnetically dominated models $\mu=1/3$,
while in the simple baryonically dominated case $\mu=1$. More general cases
involving different magnetic geometries or flows where the opening angle varies
with radius will generally lie between $1/3$ and $1$. The case of $\mu=1/3$ has
been further developed by \citet{Veres+12magnetic}, including the effects of
the shocked electrons at the external deceleration radius; the generalized
magnetic model including an arbitrary acceleration law $1/3\lesssim
\mu\lesssim1$ is considered in more detail in Veres et al. 2012 (in preparation).

{ The models typically invoked to explain the prompt emission spectral peak
fall into two broad categories. The most widely used category involves internal
shocks occurring outside the scattering photosphere. The other category
ascribes the peak to effects in or near the scattering photosphere, and within
this category the location of the photosphere and the dissipation details
depend  on whether baryonic or magnetically dominated jet dynamics are assumed.
}

The standard internal shock picture occurring outside the photosphere
\citep{Rees+94is} uses an outflow variability timescale $t_v$ { usually
tenths of} seconds and an average Lorentz factor $\eta$, for which the internal
shocks occurring at a radius $r_{\rm IS}=2~\eta^2 c t_{v}\approx
5\times10^{14}~ t_{v,-1} \eta_{2.5}^2\cm$.  With the usual magnetic field
prescription one expects an observer frame peak energy of $\ve_{\rm peak}^{\rm
IS}\approx 1~ L_{52}^{1/2} \eta_{2.5}^{-1} t_{v,-1}^{-1} \epsilon_{B,0.3}^{1/2}
\epsilon_{e,0.5}^2 \frac{2}{1+z} \MeV$, an upper limit based on a variability
$t_v=10^{-1}$ s (the light curve of GRB 110721A is rather smooth, FRED-like,
suggesting a longer variability time). A somewhat different calculation
including pair formation \citep{Guetta+01} indicates an upper limit $\ve_{\rm
peak}\lesssim 1~L_{52}^{-1/5}\epsilon_B^{1/2} \epsilon_e^{4/3}\eta_3^{4/3}
t_{v,-3}^{1/6}\frac{2}{1+z} ~\MeV$ which is more stringent. Thus, standard
internal shocks do not straightforwardly explain this burst.

In the standard photosphere model using the usual $\mu=1$ baryonic dynamics, as
pointed out by  \citet{B-Zhang+12}, the photospheres generally arise in the
coasting regime at $r>r_{\rm sat}$ where $\Gamma\sim \eta\sim$ constant, and in
the absence of dissipation the black-body peak energy and flux representing a
putative Band spectrum peak would fall below a line which excludes the
observations of GRB 110721A.

{ Recently, more detailed photospheric models have been calculated
\citep{Beloborodov10phot,Vurm+11phot} (see also also \citet{Peer+06phot,
Peer+11multicolor, Beloborodov12epeak, Vurm+12peak}), which show a peak energy
emerging from the opaque  parts of jets, through Comptonization of a thermal or
synchrotron seed component.  The numerical studies yield spectra consistent
with the observed Band shape.  These analyses generally involve dissipation
below the photosphere, and some include the effects of neutron and proton
collisions as well, assuming baryonic dynamics, with the synchrotron radiation
typically playing a sub-dominant or seed role. { \citet{Beloborodov12epeak} also showed that peak energies as high $20 \MeV$ can be obtained, explaining e.g. the high $\ve_b$ of GRB 110721A in the framework of  these models.} Other detailed models have
argued for magnetically dominated dynamics and magnetic dissipation controlling
the photospheric radiation \citep{Drenkhahn+02, Giannios+05spec,
Giannios+07photspec, Metzger+11grbmag, McKinney+11magphot}.  Along these lines,
here we propose a simple model based on magnetically dominated dynamics, where
synchrotron radiation can be expected to play a stronger role.  We calculate
the relation between the luminosity and the peak energy from synchrotron
radiation based on a photospheric dissipation model for values of $\mu$ between
the two extremes, in the context of GRB 110721A. 
}

\section{A dissipative photosphere model}

{\it Photospheric radius} We consider a dissipative photosphere, where a
substantial { fraction of the magnetic and bulk kinetic} energy is converted
into random particle energy. An extra spectral component coming from the
interaction of the photospheric photons with shock-accelerated electrons at the
external deceleration radius is neglected here, as this produces a GeV range
contribution, whereas 110721A has only a tentative GeV component.  The
photosphere arises  at optical depth $\tau=1$, in a jet which initially
accelerates as: $\Gamma=(r/r_0)^\mu$.   Here $r_0$ is launching radius at the
base of the jet. The value of the photospheric radius can be expressed in terms
of a critical Lorentz factor \citep{Meszaros+00phot} which, generalized to an
arbitrary $\mu$,  is given by 
\begin{equation}
\eta_T=(L\sigma_T/8\pi
m_p c^3 r_0)^{\mu/1+3\mu}.
\end{equation}
For $\eta>\eta_T$, the photosphere occurs in the accelerating phase, $r_{\rm
ph}=\eta_T^{1/\mu} (\eta_T/\eta)^{1/(1+2\mu)}$, while for $\eta<\eta_T$, it is
$r_{\rm ph}=\eta_T^{1/\mu}(\eta_T/\eta)^3$. The saturation radius is $r_{\rm
sat}=r_0\eta^{1/\mu}$.

The photosphere occurs in the acceleration phase if $\eta>\eta_T$, which is
typical for a magnetically dominated ($\mu=1/3$) case, where $\eta_T\simeq
120~L_{53}^{1/6} r_{0,7}^{-1/6}$. On the other hand, the photosphere is in the
coasting phase for $\eta<\eta_T$, which is typical for baryonic cases
($\mu=1$), where $\eta_T\simeq  1300~L_{53}^{1/4} r_{0,7}^{-1/4}$.  The
photospheric radius can be increased by a factor of $\sim$ few by the presence
of pairs { \citep{Beloborodov10phot, Vurm+11phot, Bosnjak+12delay,
Veres+12magnetic}.} { Also, some of the mentioned models
\citep[e.g.][]{Beloborodov10phot,McKinney+11magphot} involve a wider
dissipation region than the more narrow range around $r_{\rm ph}$. }

{\it Spectral peak energy}
The fraction of the total luminosity dissipated and converted into radiation
close to the photosphere {  depends on various uncertain model parameters, a
rough estimate being $\sim 0.5$ \citep[{ as for example
in}][]{Giannios12peak, Metzger+11grbmag, McKinney+11magphot}. { Calculations of}
the radiation from magnetic dissipation in the photosphere proceed mainly along
phenomenological lines, due to the complicated nature of the process and the
uncertainties. However, reconnection is expected to lead to MHD turbulence, and
relativistic turbulent bulk random motions rapidly become semi-relativistic
\citep{Zhang+09turb}. Semi-relativistic shocks with relative Lorentz factor
$\Gamma_r\sim 1$ can be expected from such turbulent motions.\footnote{If the
flow is baryonically dominated, where the magnetic fields are sub-dominant, one
can in principle also envisage shock dissipation near the photosphere resulting
from other effects, such as variable outflows with a wide range of variability,
recollimation, etc.}
This provides a simple alternative model for the radiation production, based on
the Fermi acceleration of electrons in these photospheric shocks}\footnote{
Note that, for the same typical magnetic field, the typical acceleration time
from reconnection and from Fermi acceleration lead formally to similar
expressions, e.g. \cite{Giannios10crmag}.  We emphasize that these are not the
usual internal shocks occurring well outside the photosphere. While usual
internal shocks require large relative Lorentz factors to achieve better
radiative efficiency and occur at radii $\sim10^{14} \cm$, here the radiation
from magnetic dissipation arises at smaller radii, and the magnetic dissipation
efficiency leading to turbulence and radiation is variously estimated to be
high by the authors cited above.}.  The minimum random Lorentz factor of the
electrons resulting from these photospheric shocks is $\gamma_{e,ph}\simeq
\epsilon_e (m_p/m_e) \Gamma_{r,0}$, where $\epsilon_e$ { $\approx 1/3$ is
the} fraction of the energy in electrons.
The synchrotron peak energy from the electrons accelerated in such shocks is
$\ve_{\rm peak}=({3q_e B'_{\rm ph}}/{4 \pi m_e c}) \gamma_{\rm e,ph}^2
({\Gamma_{\rm ph}}/{1+z})$, where ${B'}_{\rm ph}=(32 \pi \epsilon_{B} m_p c^2
{n'}_b)^{1/2} \Gamma_r$.    The physical constants have the usual meaning.
$\epsilon_B \lesssim 1$ is the fraction of the energy in magnetic form,
$n_b'=L/4\pi r_{\rm ph}^2 m_p c^3 \Gamma_{\rm ph} \eta$ is the comoving baryon
density. This latter quantity scales as $r^{-2-\mu}$ up to the saturation
radius, and as $r^{-2}$ in the coasting phase.  The Lorentz factor of the
photosphere is $\Gamma_{\rm ph}=(r_{\rm ph}/r_0)^{\mu}$ if the photosphere
occurs in the acceleration phase, and it is $\eta$ otherwise.  This results in
a peak energy which is in the MeV range (see \S \ref{sec:models}), which is the
same range as obtained from phenomenological magnetic dissipation heating and
cooling estimates by, e.g. \cite{Giannios+05spec, Giannios+07photspec} etc. The
dependence on the parameters of this peak energy is
\begin{eqnarray}
\ve_{\rm peak} \propto \left\{
\begin{array}{ll}
 L^{\frac{3\mu-1}{4\mu+2}} \eta^{-\frac{3\mu-1}{4\mu+2}}
 r_0^{\frac{-5\mu}{4\mu+2}} \epsilon_e^2 \Gamma_r^3/(1+z)		&	{\rm if~ }
 \eta>\eta_T	\\
 L^{-1/2} \eta^{3} \epsilon_e^2 \Gamma_r^3/(1+z)		&	{\rm if~ }
 \eta<\eta_T.
\end{array}
\right.
\label{eq:ebreak}
\end{eqnarray}
We provide exact values for GRB 110721A for the representative cases discussed
here in \S \ref{sec:models}.

{ The spectral indices below and above the peak are affected by various
factors.  The low energy slope can be affected by synchrotron self-absorption,
which in magnetic photospheres can be in the tens of keV range (e.g.
\cite{Giannios+05spec}).  Recent studies \citep{Sakamoto+11bat2} also indicate
that constraints on synchrotron  low energy slopes appear in only 10\% of
bursts and are less stringent than in previous studies. Above the peak, pairs
can have an effect.  Pairs created by $\gamma\gamma$ processes depend
significantly on the photon index of the spectrum above the peak energy. The
result of pair formation is a steepening break in the original spectrum.  The
photon index in the interval with the extreme peak energy is rather steep
($\beta=-3.5$) which points to an increase in the photospheric radius due to
pair creation, which is not significant, consistent with our estimate of at
most few.  If a pair component is present, which modifies the photospheric
radius by $a\simeq$ few, the peak energy decreases by $a^{-\mu/2+1}$ which
would alter the peak energy, but not in a measure to change our conclusions. A
model closely related to ours ascribes the prompt emission to magnetic
turbulence \citep{thompson94}, which results in a natural cutoff in the
spectrum at $\Gamma_{\rm ph} m_e c^2/(1+z)$. In this case we would expect only
minor effects from pair production. }

{\it Thermal peak}
Besides the synchrotron component we expect also a (modified) thermal component
from the photosphere. This component is  observed as a  thermal peak  and it is
advected from the launching radius, and cooled according to the generalized
dynamics.  We present concrete values for the temperature in Section
\ref{sec:models}, here we only show the general dependence on the parameters: %
\begin{eqnarray}
T_{\rm obs}(r_{\rm ph}) \propto \left\{
\begin{array}{ll}
 L^{\frac{14\mu-5}{12(2\mu+1)}} \eta^{\frac{2-2\mu}{6\mu+3}}
 r_0^{-\frac{10\mu-1}{6(2\mu+1)}} /(1+z)		&	{\rm if~ }
 \eta>\eta_T\\
 L^{-5/12} \eta^{8/3} r_0^{1/6} /(1+z) 	&	{\rm if~ }
 \eta<\eta_T.
\end{array}
\right.
\end{eqnarray}
There were hints of a separate thermal component below the Band peak
\citep{Page+11Xthermal, Guiriec+11thermal, Zhang+11-latgrb, mcglynn+12phot} in
previous bursts, and there are  reportedly significant thermal components also
in GRB 110721A \citep{Fermi+12epeak}.

\section{Extreme models}
\label{sec:models}

Here we consider the $\ve_{\rm peak}-L$ pairs for different $\mu$ values to
show that in the framework of a dissipative photosphere, the high peak energy
of GRB110721A can be obtained in a straightforward manner.

\begin{figure}[htbp]
\begin{center}
\includegraphics[width=0.99\columnwidth]{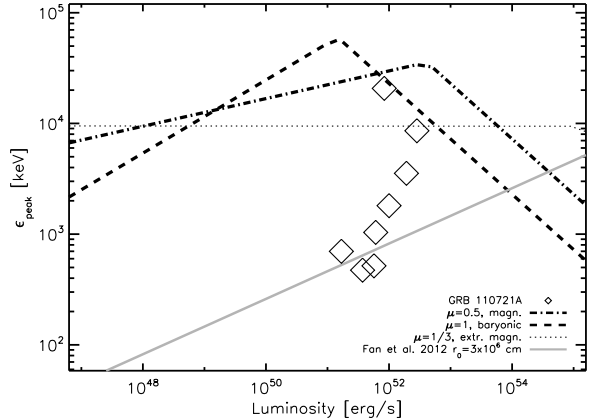}
\caption{GRB 110721A values of spectral peak energy versus peak luminosity for
different time bins advancing from top to bottom (diamonds, from
\citet{B-Zhang+12}), compared to expected limiting values from dissipative
synchrotron photosphere models (dashed: baryonic $\mu=1$, dotted: extreme
magnetic $\mu=1/3$, dot-dashed: moderate magnetic  $\mu=0.5$ (this work)). These
lines are an upper limit: values of $\ve_{\rm peak}$ and $L$ below them are
allowed. Thus, moderate magnetic and baryonic dissipative photospheres are
compatible with the highest peak energy values observed, and even the extreme
magnetic case is within a standard deviation.   The parameters used are:
for $\mu=1/3$:	$r_0=10^7 \cm$, $\epsilon_B=1$,   $\Gamma_r=2$, for $\mu=0.5$:
$r_0=10^7 \cm$, $\epsilon_B=1$,   $\Gamma_r=1.2$, for $\mu=1$:	$r_0=10^8
\cm$, $\epsilon_B=0.1$, $\Gamma_r=1.2$, and $\eta=300$  throughout. The solid
gray line is a standard (non-dissipative) photosphere curve following
\citet{Fan+12corr}, confirming the conclusion of \citet{B-Zhang+12} that
non-dissipative photospheres cannot explain the highest peak energies.	}
\label{fig:corr} \end{center} \end{figure}

\begin{figure}[htbp]
\begin{center}
\includegraphics[width=0.999\columnwidth]{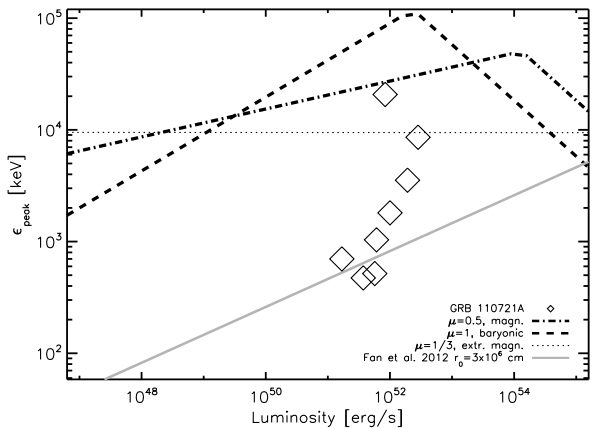}
\caption{Same notation and parameters as Figure \ref{fig:corr}, except for the coasting
Lorentz factor, which is $\eta=600$ here.
} \label{fig:corrb} \end{center}
\end{figure}

From Equation \ref{eq:ebreak}, $\eta>\eta_T$ case (photosphere in the
acceleration phase), one sees that higher luminosities, and lower Lorentz
factors and launching radii, will result in larger peak energies. Increasing
$L$ and decreasing $r_0$ will increase $\eta_T$ eventually leading to the
$\eta<\eta_T$ case (photosphere in the coasting phase).  This transition is
shown by the break in the evolution of $\ve_{\rm peak}$ with luminosity (see
Figures \ref{fig:corr} and \ref{fig:corrb} and note that in the extreme
magnetic case, the peak energy does not depend on the luminosity).

Figures \ref{fig:corr}  and \ref{fig:corrb} show that the observed highest peak
energy- luminosity pair for GRB 110721A is in the admissible region of the
diagrams, and for other reasonable parameters the peak energy is comfortably
within the operating range of these models (admissible regions are under the
modeled broken power laws, represented by dotted, dashed and dash-dotted
lines).

Here we present cases to illustrate the peak energy dependence of the
luminosity, and show that  within dissipative synchrotron photospheric models
we can reproduce the high peak energy. We have taken a redshift 
$z=0.382$\footnote{ Some observations indicate a redshift of $z\approx3.5$ for
this burst. In this case, in order to incorporate the high peak energy we need
a slightly higher  $\Gamma_r\lesssim3$, which is still not unrealistic.  }
\citep{Berger11-110721Aredshift} for the calculations presented here.  The
luminosity values in the time bins are scattered around $L=10^{52}$ erg/s (see
e.g. Figure \ref{fig:corr}) thus we take this as a reference value.

\subsection{Baryonic photosphere model ($\mu=1$)} In this model, the magnetic
field is subdominant, the dynamics are governed by the baryons in the outflow.
The critical Lorentz factor is $\eta_T\approx 740~ L_{52}^{1/4}
r_{0,7}^{-1/4}$, which puts the photosphere in the coasting phase for
moderately high $\eta\lesssim600$ values.  The peak energy will become:
$\ve_{\rm peak}\approx 16~L_{52}^{-1/2} \eta_{2.5}^{3} \Gamma_{r,0.1}^3
\epsilon_{B,-1}^{1/2}\MeV$.  This is the right order of magnitude for not too
extreme parameter values.  Lending further support for this model (the baryonic
variant) is the temperature of the thermal component, which also turns out the
right order of magnitude, $T\approx { 70}~L_{52}^{-5/12}
\eta_{2.6}^{8/3}r_{0,7}^{1/6}~\kev $.

\subsection{Extreme magnetic model ($\mu=1/3$)}  In this extreme case the break
energy does not depend on the luminosity, nor the coasting Lorentz factor.
Still, for reasonable values, we can get (see equation \ref{eq:ebreak}) a peak
value $\ve_{\rm peak}\approx 13 ~r_{0,7}^{-1/2} \Gamma_{r,0.5}^3
\epsilon_{B,0}^{1/2} \MeV $. {The temperature of the blackbody radiation is
$T\approx 3~L_{52}^{-1/60} \eta_{2.5}^{4/15}r_{0,7}^{-7/30}~ \kev $. The
dependence on the parameters is weak and this model cannot account for the
observed temperature of the black-body component. }

\subsection{ Moderate magnetic model ($\mu=0.5$)} Here magnetic fields are
dominant, but the role of baryons is more marked than in the extreme magnetic
case.  The peak energy in this intermediate variant of the model is $\ve_{\rm
peak}\approx 14~L_{52}^{1/8} \eta_{2.5}^{-1/8}r_{0,7}^{-5/8} \Gamma_{r,0.1}^3
\epsilon_{B,0}^{1/2}~\MeV$.  Again, this model can incorporate the rather large
peak energy for not too exceptional parameters. The temperature of the
black-body is lower than in the baryonic  photosphere case, $T\approx
{24}~L_{52}^{1/12} \eta_{2.5}^{1/6}r_{0,7}^{-1/3}~\kev$, but might still be
consistent with the observations.

\section{Discussion}
\label{sec:disc}
We have considered the high peak energy values of the prompt spectrum of GRB
110721A, which reach as high as $\ve_{\rm peak}\sim 15$ MeV
\citep{Fermi+12epeak}. A consideration of the usual internal shock prompt
emission spectrum model shows that such high values are unlikely in this model.
Furthermore, we confirm the conclusion of \citet{B-Zhang+12} that a
(non-dissipative) standard black-body baryonic photosphere model also cannot
explain such high peak values and fluxes. However, we show that dissipative
photospheric models, with a typical peak energy due to synchrotron radiation,
are able to accommodate such high peak energies and flux values, with
reasonable parameters, for cases where the dynamics is either baryonically or
magnetically dominated.  If the temperature of the putatively observed
black-body component can be used as a discriminant, this would seem to favor a
more baryon dominated dissipative photosphere model, $\mu$ closer to $1$,
although a moderate magnetically dominated photosphere may also be possible.

PV and PM acknowledge NASA NNX09AL40G and OTKA grant K077795 for partial
support, BBZ acknowledges the support from NASA SAO SV4-74018.  We thank Bing
Zhang, Rui-Jing Lu, En-Wei Liang and Xue-Feng Wu for supplying the time binned
data and for discussing their results { and the referee for a thorough report. }



\end{document}